\documentclass[sn-mathphys,Numbered]{sn-jnl}% Math and Physical Sciences Reference Style
\usepackage{graphicx}%
\usepackage{multirow}%
\usepackage{amsmath,amssymb,amsfonts}%
\usepackage{amsthm}%
\usepackage{mathrsfs}%
\usepackage[title]{appendix}%
\usepackage{textcomp}%
\usepackage{manyfoot}%
\usepackage{booktabs}%
\usepackage{listings}%
\usepackage{amsfonts,amssymb,stmaryrd,latexsym,amsmath,braket}
\usepackage{graphicx}
\usepackage{comment}
\usepackage{times}
\usepackage{slashed}
\usepackage[dvipsnames,usenames]{xcolor}
\usepackage{tikz}
\usetikzlibrary{quantikz}
\usepackage{lipsum,adjustbox}
\usepackage{mathtools}
\usetikzlibrary{arrows.meta}
\begin{document}

\title{Demonstration of the Rodeo Algorithm on a Quantum Computer}


\author[1]{\fnm{Zhengrong} \sur{Qian}}\email{qianzhe3@msu.edu}

\author[1]{\fnm{Jacob} \sur{Watkins}}\email{watkins@frib.msu.edu}
\author[1]{\fnm{Gabriel} \sur{Given}}\email{given@frib.msu.edu}

\author[1]{\fnm{Joey} \sur{Bonitati}}\email{bonitati@frib.msu.edu}

\author[2]{\fnm{Kenneth} \sur{Choi}}\email{kennethkchoi22@gmail.com}

\author*[1]{\fnm{Dean} \sur{Lee}}\email{leed@frib.msu.edu}

\affil[1]{\orgdiv{Facility for Rare Isotope Beams and Department of Physics and Astronomy}, \orgname{Michigan State University}, \city{East Lansing}, \state{State}, \postcode{48824}, \country{U.S.A}}

\affil[2]{\orgdiv{Department of Electrical Engineering and Computer Science}, \orgname{Massachusetts Institute of Technology}, \city{Cambridge}, \postcode{02139}, \state{Massachusetts}, \country{U.S.A}}


\abstract{The rodeo algorithm is an efficient algorithm for eigenstate preparation and eigenvalue estimation for any observable on a quantum computer. This makes it a promising tool for studying the spectrum and structure of atomic nuclei as well as other fields of quantum many-body physics.  The only requirement is that the initial state has sufficient overlap probability with the desired eigenstate. While it is exponentially faster than well-known algorithms such as phase estimation and adiabatic evolution for eigenstate preparation, it has yet to be implemented on an actual quantum device.  In this work, we apply the rodeo algorithm to determine the energy levels of a random one-qubit Hamiltonian, resulting in a relative error of $0.08\%$ using mid-circuit measurements on the IBM Q device Casablanca.  This surpasses the accuracy of directly-prepared eigenvector expectation values using the same quantum device.  We take advantage of the high-accuracy energy determination and use the Hellmann-Feynman theorem to compute eigenvector expectation values for a different random one-qubit observable.  For the Hellmann-Feynman calculations, we find a relative error of $0.7\%$. We conclude by discussing possible future applications of the rodeo algorithm for multi-qubit Hamiltonians. }


\maketitle

\section{Introduction}
Determining the eigenvalues and eigenvectors of a quantum Hamiltonian is one of the grand challenges of quantum many-body theory, where the dimensionality of the Hilbert space grows exponentially with the number of constituent particles.  In nuclear physics, there is much interest in predicting the spectrum, structure, and properties of the low-lying energy states of atomic nuclei from {\it ab initio} calculations \cite{Barbieri:2016uib,Lonardoni:2017hgs,Wirth:2017bpw,Lonardoni:2018nob,Piarulli:2017dwd,Tichai:2018vjc,Hupin:2018biv,Sun:2018fmu,Dytrych:2018vkl,Smirnova:2019yiq,Holt:2019gmc,Dawkins:2019vcr,Idini:2019hkq,Yao:2019rck,Dreyfuss:2020lss,Tichai:2020dna,Stroberg:2019mxo,Jiang:2020the,Shen:2022bak,Elhatisari:2022qfr}. Also of great interest are spectral response functions, which measure the overlap of energy states with a particular initial state such as the ground state struck by an external probe \cite{Roggero:2018hrn,Roggero:2020qoz,Raghavan:2020bze,Sobczyk:2021dwm}.

Quantum computing offers the potential for new and efficient methods for eigenvalue estimation,  eigenstate preparation, and spectral response function calculations. There are several well-known algorithms that measure energy eigenvalues by means of time evolution controlled by an auxiliary register of qubits. Some examples include quantum phase estimation \cite{Abrams:1999,Cleve:1998} and iterative quantum phase estimation \cite{Kitaev:1995qy,Svore:2013}. The rodeo algorithm (RA) is another recently introduced method that uses time evolution controlled by an auxiliary register of qubits \cite{Choi:2020pdg}. However, it is a stochastic algorithm that uses destructive interference to suppress eigenvectors with eigenvalues different from the desired target energy.   It was shown to be exponentially faster than quantum phase estimation and adiabatic evolution \cite{Farhi:2000a,Wiebe:2011a} for preparing eigenstates \cite{Choi:2020pdg}.  While several promising examples were considered, no implementation had been demonstrated on a quantum device.   

In this letter we present the first application of the RA on a quantum device and demonstrate its performance in computing the eigenvalues of a random one-qubit Hamiltonian, using the cloud-based IBM quantum computer Casablanca.  In addition to finding the energy spectrum of the Hamiltonian, we also use the Hellmann-Feynman theorem \cite{Feynman:1939} to compute the eigenvector expectation values for a different random one-qubit observable.  We note that a number of other algorithms have been recently proposed that also determine the spectrum of a Hamiltonian operator \cite{Roggero:2020qoz,Roggero:2020sgd,Sobczyk:2021ejs,Cortes:2021,Kiss:2024foh}.  There have been several reviews on quantum computing algorithms for quantum many-body systems \cite{Ayral:2023ron,Lee:2023izc} as well as new applications and improvements to the rodeo algorithm \cite{Rocha:2023bpe,Cohen:2023rhd}.  The Hellmann-Feynman theorem has also been discussed in the context of molecular forces and quantum chemistry \cite{OBrien:2021hgm,Lai:2023dga}.

\section{Applications to one-qubit Hamiltonians}
\begin{figure}[!ht]
\centering
\begin{adjustbox}{width=8.5cm}
\begin{quantikz}[transparent,slice style=blue]
	 	\lstick{object:}
	    &\qw & \gate{\exp(-iH_{\rm obj}t_k)}\gategroup[2,steps=2,style={dashed, rounded corners,fill=orange!30, inner xsep=2pt},background,label style={label position=below,anchor= north,yshift=-0.2cm}]{{Operations dependent on $t_k$}} & \qw  & \qw &\qw\rstick{} \\
	 	\lstick{\rm ancilla:}   &\gate{{\rm H}}  &\ctrl{-1} & \gate{{\rm P}(Et_k)} & \gate{{\rm H}} & \meter{}\
\end{quantikz}
\end{adjustbox}

\caption{\textbf{One cycle of the RA.}  A single cycle of the RA for a one-qubit object Hamiltonian. The object qubit is initialized to some general state $\ket{\psi_0}$, while the ancilla starts at $\ket{0}$. The boxed quantum gates depend on the random variables $t_k$, which are sampled from a Gaussian distribution of width $\sigma$. The parameter $E$ in the phase gate effectively shifts $H_{\rm obj}$ by a constant, an important element of the algorithm.}
\label{singlerun}
\end{figure}
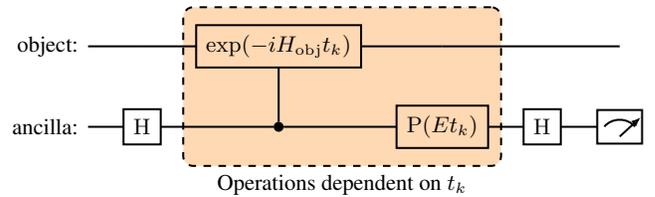

We consider a quantum register composed of two qubits.  The first qubit is our system under study, which we term the ``object" system. The second qubit is the ancilla, or ``arena'' qubit.  Any Hamiltonian $H_{\rm obj}$ on a single qubit has the form $c_I I + c_X X + c_Y Y + c_Z Z$. Here $I$ is the identity; $X,Y,Z$ are the Pauli operators; and $c_I,c_X,c_Y,c_Z$ are real coefficients.  The object system is initialized in some state $\ket{\psi_I}$, while the ancilla is initialized in the state $\ket{0}$.  We will apply $N$ successive cycles of the RA with target energy $E$, and the elements of the $k^{\rm th}$ cycle are described as follows.  The ancilla is first transformed by a Hadamard gate H.  This is followed by the controlled time evolution of $H_{\rm obj}$ for time duration $t_k$.  We then perform a phase rotation P$(Et_k)$ on the ancilla qubit, apply another Hadamard gate H, and measure the ancilla.   Mid-circuit measurements were recently enabled by IBM Q, and they allow us to reuse the same ancilla qubit for all cycles.  The elements of the cycle are illustrated in Fig.~\ref{singlerun}.

The time $t_k$ for each cycle is a random number sampled from a Gaussian distribution centered at zero and specified root-mean-square value $\sigma$.  We define a successful outcome to be the case where all $N$ successive measurements of the ancilla qubit are $\ket{0}$, which we denote compactly as the zero string $0^N$.  Suppose now that our initial state $\ket{\psi_I}$ is an eigenstate of $H_{\rm obj}$ with eigenvalue of $E_{\rm obj}$. For target energy $E$ and $N$ rodeo cycle times $\{t_k\}$ with $k = 1, \cdots, N$, the probability of a successful outcome will be denoted as $P_{0^N}(E|\{t_k\})$.  We find that
\begin{align}\label{Probs}
P_{0^N}(E|\{t_k\}) = \prod_{k=1}^N\cos^2\left[\frac{t_k}{2}(E_{\rm obj}-E)\right].
\end{align}
If we marginalize or average over all possible time values $\{t_k\}$, we find that
\begin{align}\label{ExpeProbs}
    P_{0^N}(E) = \left[\frac{1+e^{-(E_{\rm obj}-E)^2\sigma^2/2}}{2}\right]^N.
\end{align}
We see that the probability of success decays exponentially fast for $E_{\rm obj}$ not equal to $E$, while there is no loss of probability when $E_{\rm obj}$ equals $E$. Moreover, the width of this exponential is defined by $\Gamma \equiv 1/\sigma$, which has units of energy. Hence, we see that changing $\sigma$ effectively changes the magnification of our energy sensor.

In this work we are initializing the ancilla in the $\ket{0}$ state and successful measurements correspond to the ancilla in the $\ket{0}$ state.  This is a change from Ref.~\cite{Choi:2020pdg} motivated by the empirical fact that the probability to flip from $\ket{0}$ to $\ket{1}$ due to noise is much smaller than the probability to flip from $\ket{1}$ to $\ket{0}$.  With the new scheme, a flip from $\ket{1}$ to $\ket{0}$ produces a small increase in the background noise level when $E_{\rm obj}$ is different from $E$.  This is preferable to the original scheme, where a flip from $\ket{1}$ to $\ket{0}$ would instead produce a decrease in the success probability when $E_{\rm obj}$ equals $E$.

\section{Determining the energy spectrum}

\begin{figure}
\centering
\includegraphics[width=8.5cm]{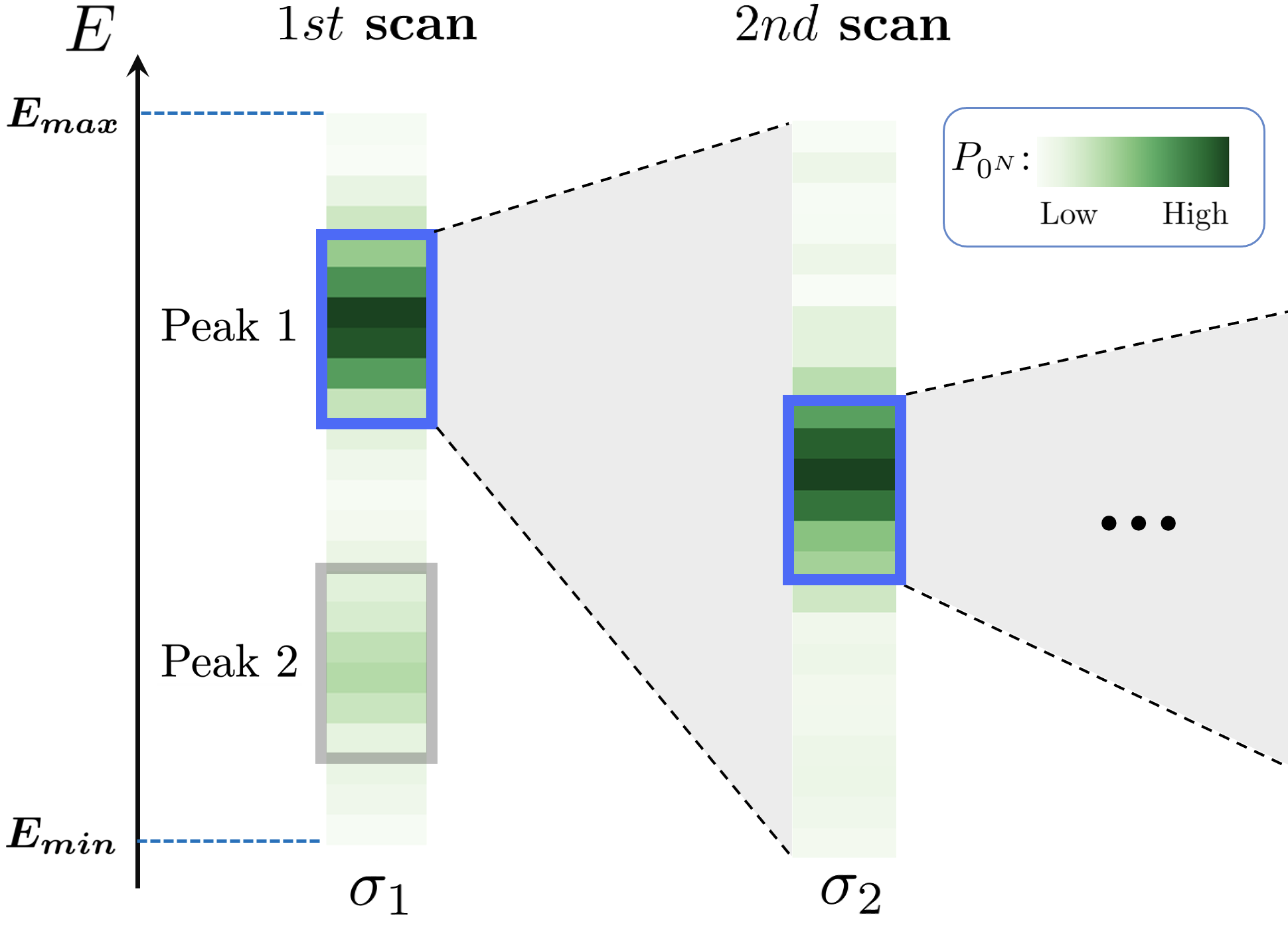}
\caption{\textbf{Sequential scans of the energy.} Each bin represents a distinct RA circuit for target energy $E$ and width parameter $\sigma$. The color and shading indicates the success probability $P_{0^N}(E)$.  Centered around each of the peaks from the first scan, a second scan is performed using with a large value of $\sigma$ and better energy resolution. This is then repeated for the third scan. }
\label{RAScan}
\end{figure}

To determine the energy eigenvalues of $H_{\rm obj}$, we implement the RA repeatedly with the target energy $E$ scanning over the energy domain from $E_{\min}$ to $E_{\max}$.  We deduce reasonable values for $E_{\min}$ and $E_{\max}$ from an estimate of the operator norm of $H_{\rm obj}$.  The energy eigenvalues will appear as peaks in the success probability distribution, $P_{0^N}(E)$.  We first locate the peaks at low resolution and then enhance the quality with finer resolution scans.  The sharpness of the energy resolution is inversely proportional to the width parameter $\sigma$.  This is illustrated in Fig.~\ref{RAScan}.   Centered around each of the peaks from the first scan, we perform a second scan with better energy resolution.  This is then repeated for the third scan.

For the calculations presented here, we will consider a one-parameter family of one-qubit Hamiltonians.  
\begin{align}
& H_{\rm obj}(\phi)=H^{(0)}+\phi H^{(1)},
\end{align}
where $H^{(0)}$ is
\begin{align}
-0.08496 I - 0.89134 X + 0.26536 Y + 0.57205 Z, 
\end{align}
and $H^{(1)}$ is
\begin{align}
-0.84537 I + 0.00673 X - 0.29354 Y + 0.18477 Z. 
\end{align}
Each of these eight coefficients were chosen as random numbers uniformly distributed from $-1$ to $1$.  We also take $N = 3$ for all the quantum circuits in this work, where $N$ is the number of successive rodeo cycle measurements.  Each cycle requires two CNOT gates, and so the total number of CNOT gates for the entire circuit is six.

The results for $H_{\rm obj}(0)$ are shown in Fig.~\ref{scanH}.  The details of the circuits used are described in the Supplemental Materials.   We take the initial object state to be $\ket{0}$, and perform three energy scans with $\sigma$ values $2, 7,$ and $12$.  We show the results obtained on the IBM Q device Casablanca, using the two connected qubits with low real-time error rate.  The results shown correspond to raw data without error mitigation.  The dashed lines indicate the expected results computed from classical calculations of the success probability.  For the first scan, we also show the results obtained using a noiseless simulator of the quantum device.  The difference between the noiseless simulator results and expected results are due to statistical errors from the finite number of measurements.
\begin{figure}
\centering
\includegraphics[width=8.5cm]{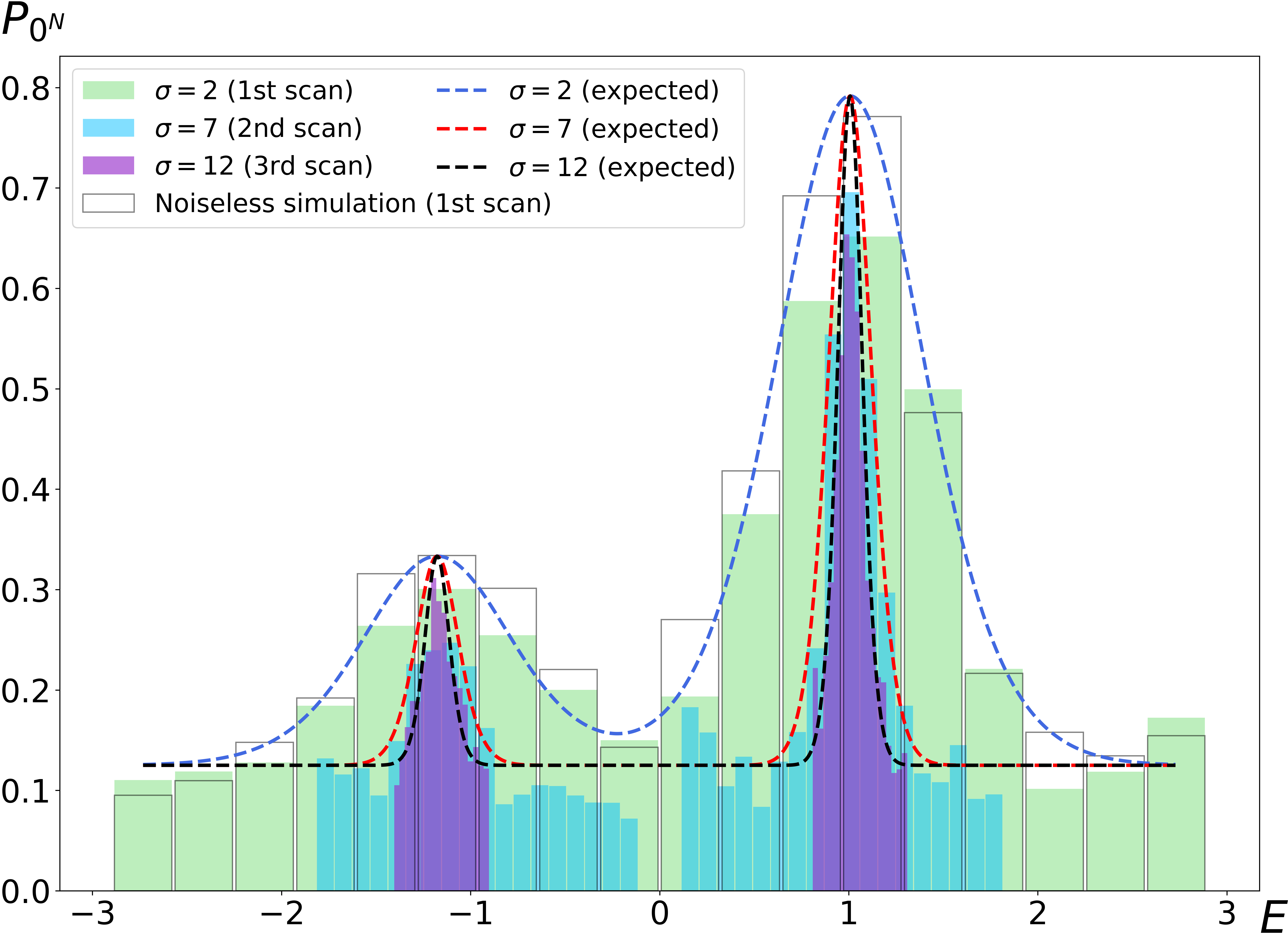}
\caption{\textbf{Energy scans for $H_{\rm obj}(0)$.} We show the results obtained on the IBM Q device Casablanca.  The dashed lines indicate the expected results computed from classical calculations of the success probability.  For the first scan, we also show the results obtained using a noiseless simulator of the quantum device.}
\label{scanH}
\end{figure}
While the noise of the quantum device reduces the peak heights below the expected values, the locations of the peaks are very well reproduced.  The peak positions are determined by fitting Gaussian functions to the success probability data near the peak.  When $E$ is more than $1/\sigma$ away from any of the eigenvalues, the probability of measuring $\ket{0}$ is approximately $1/2$ for each cycle.  With $N=3$ cycles the success probability is $(1/2)^3 = 0.125$, and this explains the background value in Fig.~\ref{scanH}.

\section{Applying the Hellmann-Feynman Theorem}
From the eigenvalues of $H_{\rm obj}(\phi)$ for small $\phi$, we can use the Hellmann-Feynman theorem to compute the expectation value of $H^{(1)}$ for the eigenstates of $H^{(0)}$ \cite{Feynman:1939}. The Hellmann-Feynman theorem is nothing more than first-order perturbation theory for the energy.  If $E_n(\phi)$ are the energy eigenvalues of $H_{\rm obj}(\phi)$ and $\ket{\psi_n(\phi)}$ the corresponding eigenstates, then we have
\begin{align}
\frac{dE_n(\phi)}{d\phi}= \braket{\psi_n(\phi)|H^{(1)}|\psi_n(\phi)}.
\end{align}
For $\phi = 0$, we get the expectation values of $H^{(1)}$ with respect to the eigenstates of $H_{\rm obj}(0)=H^{(0)}$.

\begin{figure}
    \centering
    \includegraphics[width=8.5cm]{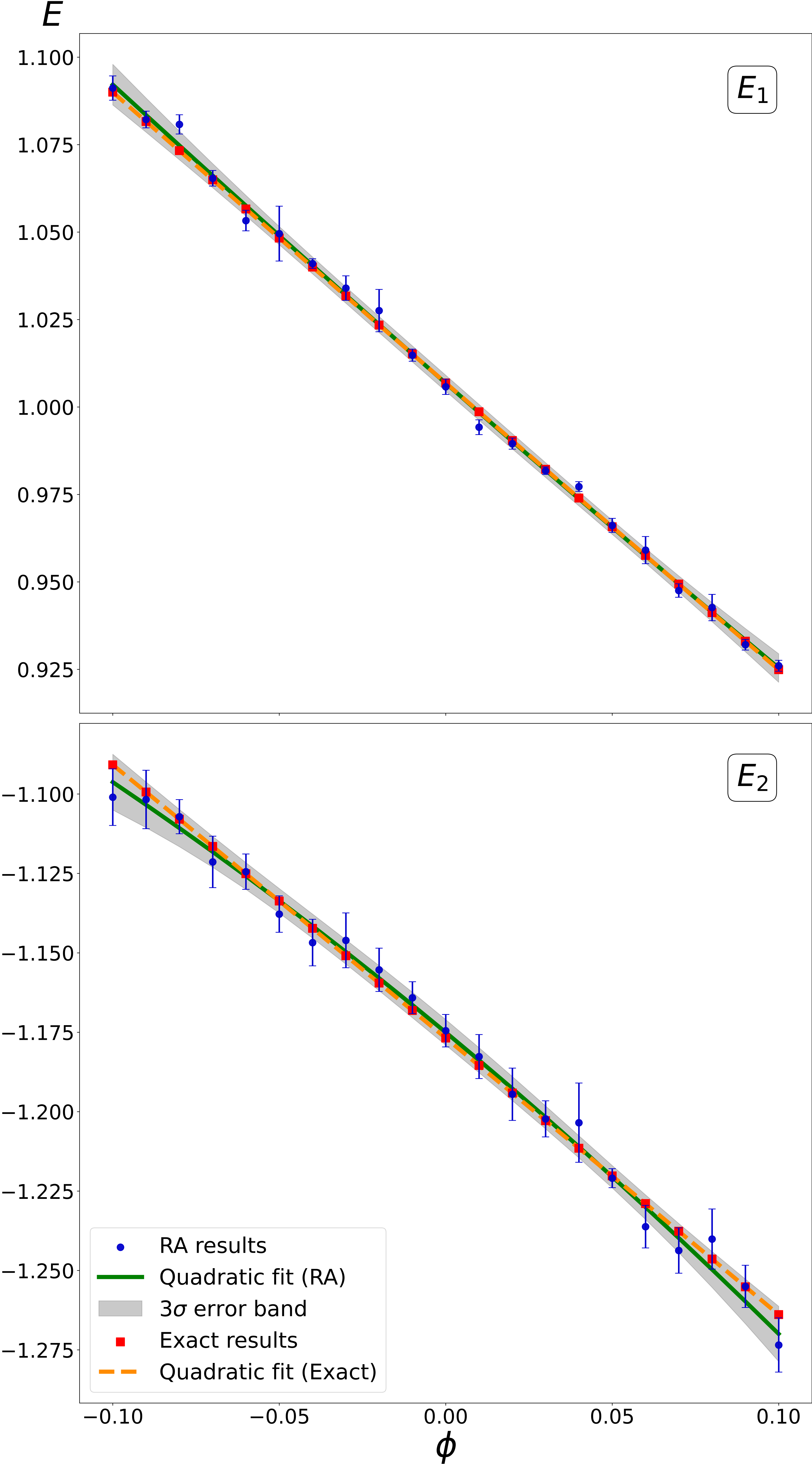}
    \caption{\textbf{Energy eigenvalues as a function of $\boldsymbol\phi$.} The top panel shows the upper eigenvalue $E_1$, and the bottom panel shows the lower eigenvalue $E_2$.  We present the RA results (filled circles), a quadratic fit to the RA results (solid line) with three-standard-deviation error bands for the fit (shaded band), exact results (filled squares), and a quadratic fit to the exact results (dashed line).}
    \label{RA_FHResults}
\end{figure}

In Fig.~\ref{RA_FHResults}, we plot the energy eigenvalues of $H_{\rm obj}(\phi)$.  The upper eigenvalue $E_1$ is shown in the top panel, and the lower eigenvalue $E_2$ is shown in the bottom panel.  Because $\phi$ is changed gradually, we only need to perform the scan for $\sigma =12$ for each additional value of E after the first. For each such value, we perform $2500$ measurements for $25$ random sets of $t_k$ for $E_1$ and $5000$ measurements for $50$ random sets of $t_k$ for $E_2$. Each set has $3$ values as $N = 3$. Plotted are the RA results (filled circles), a quadratic fit to the RA results (solid line) with three-standard-deviation error bands (shaded band), exact results (filled squares), and a quadratic fit to the exact results (dashed line).  The error bars on the individual RA data points indicate one-standard-deviation errors.  The overlap of the initial state with the upper eigenstate is somewhat larger and is the reason for the smaller error bars.

From the quadratic fit to the RA data points, we can extract energy eigenvalues of $H^{(0)}$ and the expectation values of $H^{(1)}$ with respect to the eigenstates of $H^{(0)}$.  The results are shown in Table~\ref{fit_results}. The error bars indicate the one-sigma uncertainties due to statistical noise and Gaussian peak fitting.  For comparison, we also show exact results.   From the averaged performance results for $\ket{\psi_1}$ and $\ket{\psi_2}$, the relative error in computing the energies of $H^{(0)}$ is $0.08\%$, fully consistent with our error estimates.  We note that no error mitigation is applied to these results.
In applying the Hellmann-Feynman theorem, we are measuring derivatives of the energy, and so the resulting errors are significantly larger.  Nevertheless, the uncertainty for the expectation value of $H^{(1)}$ still remains small. From the averaged performance results for $\ket{\psi_1}$ and $\ket{\psi_2}$, the relative error for the energies of $H^{(1)}$ is $0.7\%$, again fully consistent with our error estimates.

\begin{table}[h]
\centering
\begin{tabular}{|c|c|c|c|c|}
\hline 
 & $\ket{\psi_1(0)}$ & exact & $\ket{\psi_2(0)}$ & exact \\
\hline
$\braket{H^{(0)}}$ & $1.00681(66)$ & $1.00690$ & $-1.1750(12)$ & $-1.1768$  \\
\hline
$\braket{H^{(1)}}$ & $-0.8338(89)$ & $-0.8254$ & $-0.868(14)$ & $-0.8653$   \\
\hline
\end{tabular}
\caption{RA results for the energy eigenvalues of $H^{(0)}$ and the expectation values of $H^{(1)}$ with respect to the eigenstates of $H^{(0)}$.  For comparison, we also show the exact results.}
\label{fit_results}
\end{table}

\section{Prepared eigenvector expectation values}

To obtain a comparative measure of quality of the RA results, we use the same IBM Q device Casablanca to directly prepare the eigenstates $\ket{\psi_1(0)}$ and $\ket{\psi_2(0)}$ and compute the expectation values of $H^{(0)}$ and $H^{(1)}$.  When we say ``directly preparing'' the eigenstate, we mean that we apply a unitary operation that produces the desired eigenvector.  This straightforward calculation can also be viewed as an upper bound on the accuracy of the variational quantum eigensolver method for the same problem \cite{Peruzzo:2013}.

In Table~\ref{no_mitigation}, the prepared eigenvector results without measurement error mitigation are presented.  We show the expectation values of $X, Y, Z$ as well as $H^{(0)}$ and $H^{(1)}$.  The error bars are statistical errors calculated from the distribution of results obtained from 10 independent trials of 5000 measurements for each of the Pauli operators with each eigenstate.  From the averaged performance results for $\ket{\psi_1}$ and $\ket{\psi_2}$, the relative error for the expectation values of $H^{(0)}$ is $5\%$, and the relative error for the expectation values of $H^{(1)}$ is $0.6\%$.
The deviation for the expectation value of $H^{(1)}$ is smaller due to the fact that the coefficients of the Pauli matrices are, by chance, smaller for $H^{(1)}$.  But both deviations are much larger than the statistical error estimates, indicating significant systematic errors likely due to measurement bias.
\begin{table}[h]
\centering
\begin{tabular}{|c|c|c|c|c|}
\hline 
 & $\ket{\psi_1(0)}$ & exact & $\ket{\psi_2(0)}$ & exact \\
\hline
\,$\langle X\rangle$\, & -0.7455(44) & -0.8164 & 0.8055(22) & 0.8164  \\
\hline
\,$\langle Y\rangle$\, & 0.2750(36) & 0.2430 & -0.2196(25) & -0.2430  \\
\hline
\,$\langle Z\rangle$\, & 0.5356(46) & 0.5239 & -0.4632(21) & -0.5239  \\
\hline
\,$\langle H^{(0)}\rangle$\, & 0.9589(48)& 1.0069 & -1.1262(24) & -1.1768  \\
\hline
\,$\langle H^{(1)} \rangle$\, &\, -0.8321(14)\, &\, -0.8254\, &\, -0.86109(84)\, &\, -0.8653\,  \\
\hline
\end{tabular}
\caption{Prepared eigenvector results without measurement error mitigation.}
\label{no_mitigation}
\end{table}

We have therefore analyzed the same data using measurement error mitigation.  Preceding the $10$ independent trials, we collect data for the $2 \times 2$ ``calibration matrix'', which gives the probability of measuring $\ket{0}$ or $\ket{1}$ when the state is prepared in a pure $\ket{0}$ or $\ket{1}$ state. We then multiply our measurement statistics for the expectation values of the Pauli operators by the inverse of the ``calibration matrix''.  The results are presented in Table~\ref{mitigation}. We see that measurement error mitigation has removed much of the error. With averaged performance results for $\ket{\psi_1}$ and $\ket{\psi_2}$, the relative error for the expectation values of $H^{(0)}$ is now $0.2\%$ and the relative error for the expectation values of $H^{(1)}$ is $0.5\%$. However, the expectation values of the Pauli operators have residual errors that are larger than the statistical errors.  This indicates that there are remaining systematic errors, and the total error cannot be reduced significantly further by increasing the measurement statistics.  All of the runs were performed in a single day, spanning a time window no more than a few hours from the calibration time.
\begin{table}[h]
\centering
\begin{tabular}{|c|c|c|c|c|}
\hline 
 & $\ket{\psi_1(0)}$ & exact & $\ket{\psi_2(0)}$ & exact \\
\hline
\,$\langle X\rangle$\, & -0.8119(46) & -0.8164 & 0.8152(27) & 0.8164  \\
\hline
\,$\langle Y\rangle$\, & 0.2569(83) & 0.2430 & -0.2596(79) & -0.2430  \\
\hline
\,$\langle Z\rangle$\, & 0.5297(80) & 0.5239 & -0.5151(89) & -0.5239  \\
\hline
\,$\langle H^{(0)} \rangle$\, & 1.0100(65) & 1.0069 & -1.1751(60) & -1.1768  \\
\hline
\,$\langle H^{(1)}\rangle$\, &\, -0.8283(28)\, &\, -0.8254\, &\, -0.8589(29)\, &\, -0.8653\,  \\
\hline
\end{tabular}
\caption{Prepared eigenvector results with measurement error mitigation.}
\label{mitigation}
\end{table}

\section{Discussion and Outlook}
We have found that the RA for the random one-qubit Hamiltonian $H^{(0)}$ achieves a relative error of $0.08\%$ for the energy eigenvalues.  This is better than the relative error obtained for the error-mitigated expectation values of $H^{(0)}$ using the directly-prepared eigenvectors.  While the object Hamiltonian acts on only one qubit, the implementation described in this work involves six two-qubit CNOT gates.  The resulting loss of fidelity due to gate errors, measurement errors, and qubit decoherence is far higher than $0.08\%$.  The fact that the RA is delivering accurate results for the energy without any error mitigation can be attributed to its unusual and robust design.  Even in the presence of significant noise, the RA succeeds in its strategy of reducing the spectral weight of eigenstates with the wrong energy.  While the noise will result in some reduction of the spectral weight of the desired eigenstate, it is still possible, with sufficient statistics, to distinguish the signal above the random background.  By sufficient statistics, we mean that the number of measurements must be large enough so that the statistical noise of the random background is smaller than the size of the signal.

For this one-qubit benchmark calculation, one could likely reach even lower relative errors using the RA, provided that the gates of the device are calibrated with sufficiently high accuracy.  As noted in Ref.~\cite{Choi:2020pdg}, for energy eigenvalue determination with relative error $\epsilon$, the computational effort scales as $O[(\log \epsilon)^2/(p \epsilon)]$, where $p$ is the squared overlap of the initial state with the target eigenvector. We contrast this with the $O(1/\epsilon^2)$ scaling of the computational effort, due to statistical errors, for directly-prepared eigenvector expectation values.  The $O(1/\epsilon^2)$ estimate is also a lower bound for the computational scaling of variational quantum eigensolvers.  We must also add the additional computational effort required for the variational search to prepare the eigenstate with the required error tolerance.

Hellmann-Feynman theorem calculations of the expectation value $H^{(1)}$ using the RA have a relative error of $0.7\%$, or about one order of magnitude larger.  This larger error comes from the fact that we must compute numerical derivatives of $E_n(\phi)$. We nevertheless have established that the Hellmann-Feynman theorem can be used to compute expectation values of observables accurately on a quantum device.  In order to compute operator expectation values with relative error $\epsilon$, we must compute the energies of $H_{\rm obj}(\phi)$ with error tolerance $O(\epsilon^2)$ for values of $\phi$ of size $O(\epsilon)$.  So the corresponding computational scaling is $O[(\log \epsilon)^2/(p \epsilon^2)]$

In this work we have demonstrated the performance of the RA for a general one-qubit object Hamiltonian.  Due to these promising results, we are now working with collaborators to test the performance of the RA on multi-qubit object Hamiltonians.  For the multi-qubit Hamiltonian, we will have additional systematic errors arising from the need to perform a Suzuki-Trotter decomposition of the time evolution operator \cite{Trotter:1959,Suzuki:1976a,Childs:2019b}.  However, one can still define an effective Hamiltonian for the multi-qubit system that exactly reproduces the Trotterized time evolution. Based on the results presented here, there is reason to believe that the eigenvalues of the multi-qubit effective Hamiltonian can also be determined with good accuracy, provided that the signal is strong enough to distinguish above the random background.  The RA may then be used in the future to study the spectrum and structure of nuclear states as well as spectral response functions. 
 There is also potential for addressing quantum many-body problems in other fields such as condensed matter physics, quantum chemistry, and ultracold atoms and molecules. For large quantum systems, however, we are faced with the problem that the overlap of the initial state with the eigenstate of interest becomes very small.  For this reason, we are currently investigating efficient preconditioning methods to increase the overlap of the initital state with the targeted eigenstate.

\paragraph*{Acknowledgements}
We are grateful for discussions with Max Bee-Lindgren, Natalie Brown, Matt DeCross, Michael Foss-Feig, Christopher Gilbreth, David Hayes, Caleb Hicks, Brian Neyenhuis, Erik Olsen, Avik Sarkar, and Xilin Zhang.  We also thank Jay Gambetta, Sebastian Hassinger, and IBM Q for the research accounts and access to hardware that made these quantum calculations possible.  This work benefited from resources of the Oak Ridge Leadership Computing Facility, which is a
DOE Office of Science User Facility supported under Contract DE-AC05-00OR22725.  We acknowledge financial support from the U.S. Department of Energy through grants DE-SC0021152, DE-SC0013365, DE-SC0023658, DE-SC0023175, and DE-SC0024586.  We also acknowledge financial support from the U.S. National Science Foundation (NSF) through grant PHY-2310620 and the NSF Graduate Research Fellowship through grant DGE-1848739.

\bibliography{References}

\clearpage

\section*{Supplemental Material}
In this section we provide further details of the quantum circuits used in the calculations.  The one-qubit object Hamiltonian, $H_{\rm obj}$, can be written as $c_I I + c_X \sigma_X + c_Y \sigma_Y + c_Z \sigma_Z$, where $\sigma_X, \sigma_Y, \sigma_Z$ are the Pauli matrices and $I$ is the identity matrix. The slight change in Pauli matrix notation is useful for the vector index contractions to follow.  We can therefore write the time evolution operator as 
\begin{align}
    U(t)=e^{-iH_{\rm obj}t}=e^{-ic_It}e^{-\frac{i\theta}{2}\hat{n}\cdot\vec{\sigma}}\equiv e^{-ic_It}R_{\hat{n}}(\theta),
\end{align} where $R_{\hat n}(\theta)$ is a rotation matrix about the three-dimensional unit vector $\hat{n}$ by angle $\theta$ and parametrizes any matrix in the defining representation of the group $SU(2)$.  We have
\begin{align} \label{eq:Rnhat}
& R_{\hat{n}}(\theta) = e^{-\frac{i\theta}{2}\hat{n}\cdot\vec{\sigma}}\nonumber \\
&=\begin{bmatrix}
\cos(\frac{\theta}{2})-i\sin(\frac{\theta}{2})n_Z & -i\sin(\frac{\theta}{2})(n_X-in_Y)\\
-i\sin(\frac{\theta}{2})(n_X+in_Y) & \cos(\frac{\theta}{2})+i\sin(\frac{\theta}{2})n_Z
\end{bmatrix},
\end{align}
\\
where
\begin{align}
\theta = 2t\sqrt{c_X^2+c_Y^2+c_Z^2},
\end{align}
and
\begin{align}
\hat{n}=\frac{1}{\sqrt{c_X^2+c_Y^2+c_Z^2}}\begin{bmatrix}
c_X\\
c_Y\\
c_Z
\end{bmatrix}
=\begin{bmatrix}
n_X\\
n_Y\\
n_Z
\end{bmatrix}.
\end{align}
From the documentation of Qiskit, the open source software kit for IBM Q devices, a generic single-qubit quantum operation $U$ is parameterized with three Euler angles $\gamma,\beta,\delta$,
\begin{align}\label{eq:U3}
U(\gamma,\beta,\delta) &=
\begin{bmatrix} 
\cos\left(\frac{\gamma}{2}\right) & -e^{i\delta} \sin\left(\frac{\gamma}{2}\right) \\
e^{i\beta} \sin\left(\frac{\gamma}{2}\right) & e^{i(\delta + \beta)} \cos\left(\frac{\gamma}{2}\right)
\end{bmatrix}.
\end{align}
Applying the $Z-Y$ decomposition for a single qubit \cite{Nielsen:2012yss}, we can rewrite the parametrization as
\begin{align}\label{U3ZY}
U(\gamma,\beta,\delta)
&= e^{i\frac{\delta+\beta}{2}}
\begin{bmatrix}
e^{-i\frac{\delta+\beta}{2}}\cos\left(\frac{\gamma}{2}\right) & -e^{i\frac{\delta-\beta}{2}}\sin\left(\frac{\gamma}{2}\right) \nonumber \\
e^{-i\frac{\delta-\beta}{2}}\sin\left(\frac{\gamma}{2}\right) & e^{i\frac{\delta+\beta}{2}}\cos\left(\frac{\gamma}{2}\right)
\end{bmatrix} \nonumber \\
&= e^{i\frac{\delta+\beta}{2}}R_Z(\beta)R_Y(\gamma)R_Z(\delta) \nonumber \\
&\equiv e^{i\frac{\delta+\beta}{2}}R_{\hat{n}}(\theta).
\end{align}
We equate the upper-left and lower-left entries of the matrices $R_{\hat{n}}(\theta)$ and $R_Z(\beta)R_Y(\gamma)R_Z(\delta)$ and obtain the following constraints:
\begin{align} \label{ul}
\cos\left(\frac{\delta+\beta}{2}\right)\cos\left(\frac{\gamma}{2}\right) &= \cos\left(\frac{\theta}{2}\right),\\
-\sin\left(\frac{\delta+\beta}{2}\right)\cos\left(\frac{\gamma}{2}\right) &= -n_Z\sin\left(\frac{\theta}{2}\right),\\
\cos\left(\frac{\delta-\beta}{2}\right)\sin\left(\frac{\gamma}{2}\right) &= n_Y\sin\left(\frac{\theta}{2}\right),\\
-\sin\left(\frac{\delta-\beta}{2}\right)\sin\left(\frac{\gamma}{2}\right) &= -n_X\sin\left(\frac{\theta}{2}\right).
\end{align}
We remove the global phase from the definition of the $U$ gate and instead implement the  overall phase controlled by the $c_I I$ term.  This generates two extra terms in the argument of the phase gate, $\xi$.  The final results for the parameters are
\begin{align} \label{gensol}
\delta &= \tan^{-1}\left[n_Z\tan\left(\frac{\theta}{2}\right)\right]+\tan^{-1}\left(\frac{n_X}{n_Y}\right), \\
\beta &= \tan^{-1}\left[n_Z\tan\left(\frac{\theta}{2}\right)\right]-\tan^{-1}\left(\frac{n_X}{n_Y}\right), \\
\gamma &= 2\cos^{-1}\left[\frac{\cos(\theta/2)}{\cos\left(\frac{\delta+\beta}{2}\right)}\right], \\
\xi &= -c_I t-\frac{\delta+\beta}{2}.
\end{align}

\end{document}